\documentclass[12pt,preprint]{aastex}

\def\ltsima{$\; \buildrel < \over \sim \;$}
\def\gtsima{$\; \buildrel > \over \sim \;$}
\def\lsim{\lower.5ex\hbox{\ltsima}}
\def\gsim{\lower.5ex\hbox{\gtsima}}
\def\lapp{\ifmmode\stackrel{<}{_{\sim}}\else$\stackrel{<}{_{\sim}}$\fi}
\def\gapp{\ifmmode\stackrel{>}{_{\sim}}\else$\stackrel{<}{_{\sim}}$\fi}

\def\msol{\,\mathrm{M}_\odot}

\usepackage{amsmath}
\usepackage{textcomp}
\usepackage{amssymb}
\usepackage{multirow}
\usepackage{booktabs}
\usepackage{graphicx}

\newdimen\minuswidth    
\setbox0=\hbox{$-$}
\minuswidth=\wd0
\catcode`@=\active
\def@{\kern\minuswidth}
\setbox0=\hbox{\rm0}
 
\shorttitle{} 
\shortauthors{Massari et al.}
 
\begin{document} 
\title{{\tt Ceci n'est pas} a globular cluster: the metallicity distribution of the stellar system Terzan~5
\footnote{ 
Based on FLAMES observations performed at the European Southern Observatory, 
proposal numbers 087.D-0716(B), 087.D-0748(A) and 283.D-5027(A), and at the W. M. Keck Observatory. Keck 
is operated as a scientific partnership among the California Institute of 
Technology, the University of California, and the National Aeronautics and Space Administration.
The Observatory was made possible by the generous financial support of the W. M. Keck Foundation.}
}

\author{
D. Massari\altaffilmark{2},
A. Mucciarelli\altaffilmark{2},
F. R. Ferraro\altaffilmark{2},
L. Origlia\altaffilmark{3},
R. M. Rich\altaffilmark{4},
B. Lanzoni\altaffilmark{2},
E. Dalessandro\altaffilmark{2},
E. Valenti\altaffilmark{5}, 
R. Ibata\altaffilmark{6},
L. Lovisi\altaffilmark{2},
M. Bellazzini\altaffilmark{3},
D. Reitzel\altaffilmark{4}
}
\affil{\altaffilmark{2} Dipartimento di Fisica e Astronomia, Universit\`a degli Studi
di Bologna, v.le Berti Pichat 6/2, I$-$40127 Bologna, Italy}
\affil{\altaffilmark{3}INAF-Osservatorio Astronomico di Bologna, via
  Ranzani 1, 40127, Bologna, Italy} 
\affil{\altaffilmark{4} 
Department of Physics and Astronomy, Math-Sciences 8979, UCLA, Los Angeles, CA 90095-1562, USA} 
\affil{\altaffilmark{5}European Southern Observatory,
  Karl-Schwarzschild-Strasse 2, 85748 Garching bei M\"{u}nchen,
  Germany} 
\affil{\altaffilmark{6} 
Observatoire Astronomique, Universit\'e de Strasbourg, CNRS, 11, rue de l'Universit\'e. 
F-67000 Strasbourg, France}

\date{30 Jul, 2014}

\begin{abstract}

We present new determinations of the iron abundance for $220$ stars belonging to the stellar system Terzan 5 in the 
Galactic bulge. The spectra have been acquired with FLAMES at the Very Large Telescope of the European Southern 
Observatory and DEIMOS at the Keck II Telescope. This is by far the largest spectroscopic sample of stars ever observed 
in this stellar system. From this dataset, a subsample of targets with spectra unaffected by TiO bands was extracted 
and statistically decontaminated from field stars. Once combined with $34$ additional stars previously published by our 
group, a total sample of $135$ member stars covering the entire radial extent of the system has been used to determine 
the metallicity distribution function of Terzan 5.  The iron distribution
clearly shows three peaks: a super-solar component at
[Fe/H]$\simeq0.25$ dex, accounting for $\sim29$\% of the sample, a dominant sub-solar
population at [Fe/H]$\simeq-0.30$ dex, corresponding to $\sim62$\% of
the total, and a minor ($6$\%) metal-poor component at [Fe/H]$\simeq-0.8$
dex. Such a broad, multi-modal metallicity distribution demonstrates that Terzan 5
is not a genuine globular cluster but the remnant of a much more complex stellar system.

\end{abstract}
 
\keywords{stellar system: individual (Terzan~5);\ stars:\ abundances;\ techniques:\ spectroscopic}

\section{INTRODUCTION}

Terzan~5 is a stellar system located in the bulge of our Galaxy,
at a distance of 5.9 kpc (\citealt{valenti}), historically
classified as a globular cluster (GC).  Its location ($l=3.8395$\textdegree,
$b=1.6868$\textdegree) corresponds to a highly extincted region of the sky,
with an average color excess $E(B-V)=2.38$ mag (\citealt{valenti}) and
a patchy structure of dust clouds that causes the extinction to vary
spatially by $\sim0.7$ mag over a projected spatial scale of few
arcminutes (\citealt{massari}).  Because of optical observations of
Terzan~5 are extremely challenging, its true nature remained hidden
behind this dusty curtain for a long time, until high-resolution
near-infrared observations revealed the presence of two distinct
stellar populations (\citealt{f09}, hereafter F09).  Two Red
Clumps (RCs) well separated in color ($\delta$(J-K)$\sim0.2$ mag) and
magnitude ($\Delta K\sim0.3$ mag) have been detected in the
near-infrared color magnitude diagram (CMD) obtained with the
Multi-conjugate Adaptive optics Demonstrator (MAD) mounted at the Very
Large Telescope.  The analysis of near-infrared high-resolution
spectra (\citealt{origlia97}) promptly acquired with NIRSPEC (\citealt{nirspec}) 
at Keck II telescope demonstrated that both populations belong to
Terzan~5 and have very different iron abundances
\citep[F09;][hereafter O11]{origlia}: [Fe/H]$=-0.25$ dex for the
component corresponding to the faint RC, [Fe/H]$=+0.27$ dex for the
bright RC.  Recently, \citet[][hereafter O13]{o13} discovered the
presence of a third, more metal-poor component at [Fe/H]$=-0.79$.
Such a large spread in the iron abundance of Terzan 5 stars
clearly indicates that its initial mass had to be much larger than the
current one ($10^{6} \msol$, \citealt{l10}), in order to retain the 
high-velocity, iron-enriched gas ejected by supernovae (SN).
Moreover, O11 did not find any hint of the spreads and
anti-correlations among light elements commonly observed in Galactic
GCs (see \citealt{carretta10}). Instead, the $\alpha$-element
abundance patterns of the three populations (see O11, O13) turned out to be
strikingly similar to those observed in the bulge field stars (e.g., 
\citealt{ful07, hill11, rich12, johnson11, johnson12, johnson13, johnson14, ness13}), 
with $\alpha$-enhancement up to about solar [Fe/H] and a progressively
decreasing [$\alpha$/Fe] towards the solar ratio at super-solar [Fe/H].

The observational evidence collected so far suggests that Terzan~5 is
not a genuine GC, but a stellar system that experienced complex star
formation and chemical enrichment histories.  In order to accurately
reconstruct its evolution, a first crucial step is to precisely
determine the metallicity distribution of its stellar populations,
based on a statistically significant sample of stars. This
paper reports the iron abundance for a sample 
of 220 giants distributed over the entire radial
extent of Terzan 5, from the innermost regions, out to the tidal
radius. In Section \ref{obs} we present the analyzed sample. In Section \ref{atmpar}
we describe the chemical abundance analysis. The overall error budget is
discussed in Section \ref{caveats}. Finally in Section \ref{results} 
we finally present the metallicity distribution and in Section \ref{discuss} 
we draw our conclusions.

\section{OBSERVATIONS AND DATA REDUCTION}\label{obs}

This work is part of a large spectroscopic survey of stars in the direction of
Terzan 5, aimed at characterizing the kinematical and chemical properties of the stellar populations
within the system and in the surrounding Galactic bulge field. While the overall
survey will be described in a forthcoming paper (Ferraro et al. 2014 in preparation) and the 
properties of the field around Terzan 5 have
been discussed in \citet[][hereafter M14a]{m14}, here we focus on 
the metallicity distribution of Terzan 5.

\begin{figure}
\plotone{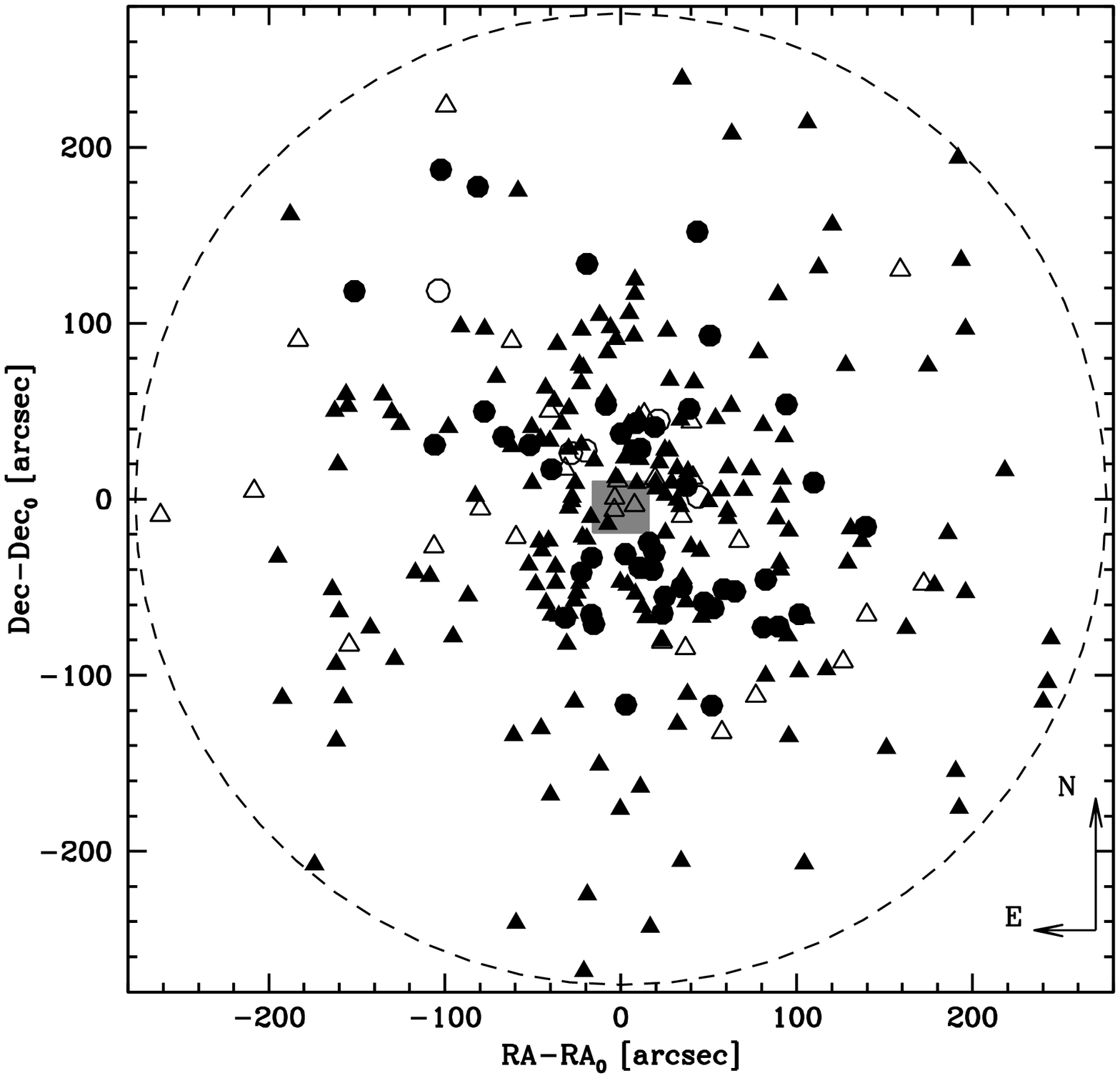}
\caption{\small Spatial distribution of the spectroscopic targets in Terzan 5. 
FLAMES and DEIMOS targets are shown as triangles and circles, respectively. 
The central gray square marks the region where the NIRSPEC targets are located
(see Section \ref{sample} for the details about the membership). 
Filled symbols mark targets for which the iron abundance was measured while 
empty symbols are used to indicate targets affected by TiO contamination for which 
no abundance determination was possible. The dashed circle
marks the tidal radius of the system, $r_t=276\arcsec =7.9$ pc ($100\arcsec$ corresponding to 2.86 pc at the distance of
Terzan 5).}
\label{map}
\end{figure}

This study is based on a sample of stars located within the tidal radius of Terzan 5 
(r$_{t}\simeq300\arcsec$; \citealt{l10, miocchi13}) observed with two different instruments:
FLAMES \citep{pasquini} at the ESO Very Large Telescope (VLT) and DEIMOS (\citealt{faber}) at the Keck II
Telescope. The spectroscopic targets have been selected from the optical photometric
catalog of Terzan~5  described in \citet{l10} along the brightest portion ($I<17$) of the
red giant branch (RGB). In order to avoid contamination from other sources, in the selection
process of the spectroscopic targets we avoided stars with bright
neighbors (I$_{neighbor}<{\rm I_{star}}+1.0$) within a distance of 2\arcsec.    
The spatial distribution of the observed targets is shown in Fig.~\ref{map}.

(1)~{\it FLAMES dataset}--- This dataset has been collected under three different programs (ID: 087.D-0716(B), PI: Ferraro, ID:
087.D-0748(A), PI: Lovisi and ID: 283.D-5027(A), PI: Ferraro). As already described in
M14a, all the spectra have been obtained using the HR21 setup in the GIRAFFE/MEDUSA mode,
providing a resolving power of R$\sim16200$ and a spectral coverage ranging from 8484 \AA{} to
9001 \AA{}.  This grating has been chosen because it includes the prominent Ca~II triplet
lines, which are widely used features for radial velocity estimates, even in low signal-to-noise
ratio (SNR) spectra.  Several metal lines (mainly of Fe~I) lie in this spectral range, thus
allowing a direct measurement of [Fe/H]. In order to reach SNR$\sim$40-50 even for the
faintest ($I\sim17$) targets, multiple exposures with integration  times ranging from 1500 s to
2400 s (depending on the magnitude of the targets) have been secured for the majority  of
the stars.  In order to reduce the acquired spectra we used the FLAMES-GIRAFFE ESO
pipeline\footnote{http://www.eso.org/sci/software/pipelines/}. This
includes bias-subtraction, flat-field correction, wavelength calibration with a
standard Th-Ar lamp, resampling at  a constant pixel-size and extraction of one-dimensional
spectra.  Because of the large number of O$_{2}$ and OH emission lines in this spectral range,
a correct sky subtraction is a primary requirement.  Thus, in each exposure $15$-$20$
fibers have been used to measure the sky. The master sky spectrum obtained as the median of
these spectra has been then subtracted from the stellar ones. Finally, all the spectra have
been reported to zero-velocity and in the case of multiple exposures they have been co-added
together. 

(2)~{\it DEIMOS dataset}--- This spectral dataset has been acquired 
by using the 1200 line/mm grating coupled  with the GG495 and GG550
order-blocking filters.  The spectra cover the $\sim$6500-9500 \AA{} wavelength range with  a
resolution of R$\sim7000$ at $\lambda\sim8500$ \AA{}.  An exposure time of $600$ s for
each pointing allowed to reach SNR$\sim50-60$ for the brightest stars and  SNR$\sim15-20$ for
the faintest ones ($I\sim17$ mag). We used the package  described in \cite{ibata11} for an
optimal reduction and extraction of the DEIMOS spectra. 

\begin{figure}
\plotone{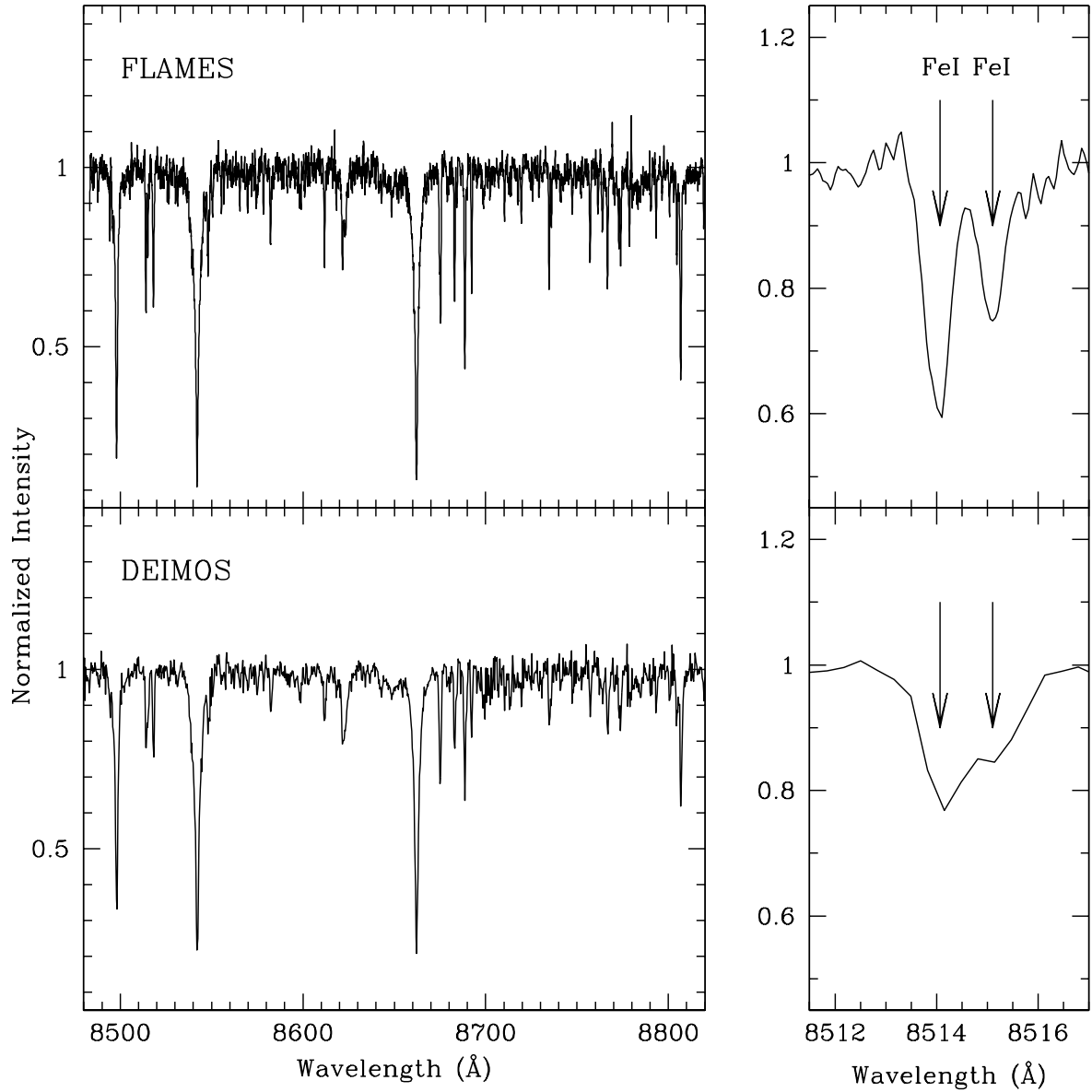}
\caption{\small The Ca II triplet spectral region for star 34, as obtained from  FLAMES (left-upper panel) and  
DEIMOS (left-lower panel) observations. The right panels show the zoomed spectra around two Fe~I lines 
used in the analysis.}
\label{spec}
\end{figure}

For sake of comparison, Fig.~\ref{spec} shows two spectra of the same star observed with 
FLAMES (top panels) and with DEIMOS (bottom panels).

\section{ANALYSIS}\label{atmpar}

\subsection{Atmospheric parameters}\label{vturb}

\begin{figure}
\plotone{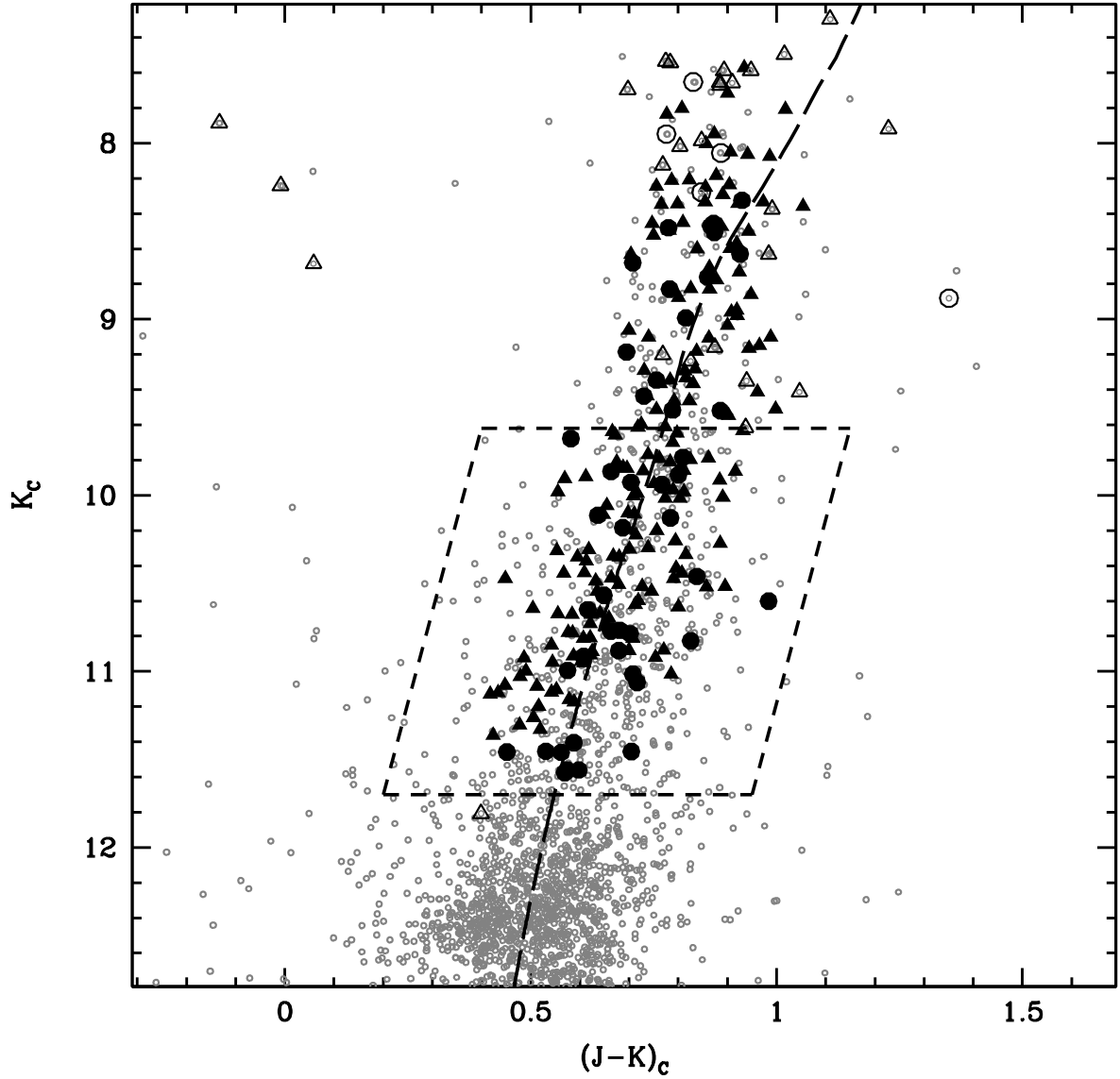}
\caption{\small Infrared CMD of Terzan~5 corrected for differential reddening. 
Symbols are as in Fig.~\ref{map}, with empty  symbols marking the targets affected
by TiO contamination.
The BaSTI isochrone with an age of 12 Gyr and metallicity Z=0.01 used to derive the atmospheric 
parameters is also shown as a long-dashed line.
The box delimited by the short-dashed line indicates the sample not affected by TiO contamination 
that was selected 
to compute  
the metallicity distribution.}
\label{cmds}
\end{figure}

Effective temperatures (T$_{{\rm eff}}$) and surface gravities ($\log$ $g$) for each target
 have been derived 
from near infrared photometry in order to minimize the effect of possible residuals  in the
differential reddening correction.  The $(K, J-K)$ CMD has been obtained by combining the SOFI
catalog of \citet{valenti} for the central  $2.5$\arcmin$ \times 2.5$\arcmin~ and 2MASS
photometry in the outermost regions. 
Magnitudes and colors of each star 
have been corrected for differential extinction according to their spatial location with respect to the 
center of Terzan 5. For stars in the innermost regions, lying within the field of view (FoV) of the 
ACS/HST observations (see \citealt{l10}), the reddening map 
published in \cite{massari}\footnote{A webtool able to compute the reddening in the
direction of Terzan 5 is freely available at the Cosmic-lab website, {\tt
http://www.cosmic-lab.eu/Cosmic-Lab/Products.html}} has been adopted. Instead  
the correction for stars in the outer regions has been estimated from
the new differential reddening map described in M14a (see their Figure 4). 
The target positions in the reddening-corrected CMD 
are shown in Fig.~\ref{cmds}. In
order to estimate T$_{{\rm eff}}$ and $\log$~$g$, the position of each target in the
reddening-corrected CMD has been projected onto a reference isochrone. 
Following F09,  we adopted a 12 Gyr-old isochrone 
extracted from the BaSTI database (\citealt{basti2}) 
with metallicity Z$=0.01$ (corresponding to [Fe/H]$=-0.25$),
$\alpha$-enhanced chemical mixture and helium content Y$=0.26$ dex  (well reproducing the dominant stellar population 
in Terzan 5, see O11). The isochrone is shown as dashed line in Fig.~\ref{cmds}.
Since Terzan~5 hosts at least two stellar populations, but they are photometrically
indistinguishable in the  near-infrared plane, in Section \ref{caveats}, we discuss the effect
of using isochrones with different metallicities and ages.

As already explained in M14a, the small number (about 10) of Fe~I lines
observed in the FLAMES and DEIMOS spectra  (see Section \ref{analysis}) prevents us from deriving a reliable 
spectroscopic determination of the microturbulent velocity  ($v_{turb}$; see \citealt{m11} for a review of the different
methods to estimate this parameter). Therefore, for homogeneity with our previous work we adopted
the same value, $v_{turb}=$1.5 km$\,$s$^{-1}$, which is a reasonable assumption for cool giant stars
(see also \citealt{zoccali08, johnson13}).

\subsection{Chemical analysis}\label{analysis}

We adopted the same Fe~I linelist and the same techniques to analyze the spectra and to determine the chemical 
abundances as those used in M14a.

(1)~{\it FLAMES data-set}---
We performed the chemical analysis using the package GALA \citep{gala}\footnote{GALA  is freely
distributed at the Cosmic-Lab project website,  {\tt http://www.cosmic-lab.eu/gala/gala.php}}, an
automatic tool to derive chemical abundances  of single, unblended lines by using their measured
equivalent widths (EWs).  The adopted model atmospheres have been calculated with the ATLAS9 code
\citep{atlas}. Following the  prescriptions by M14a, we performed the analysis
running GALA with  all the model atmosphere parameters fixed and allowing 
only the metallicity to vary  iteratively in order to match the iron
abundance measured from EWs. The latter were measured by using
the code 4DAO (\citealt{4dao})\footnote{4DAO is freely distributed at the website 
{\tt http://www.cosmic-lab.eu/4dao/4dao.php.}}.  This code runs DAOSPEC \citep{daospec} for
large sets of spectra, tuning automatically the main input parameters used by DAOSPEC. It also
provides graphical outputs  that are fundamental to visually check the quality of the fit for each
individual spectral line. EW errors are estimated by DAOSPEC as the standard deviation of the local
flux residuals \citep[see][]{daospec}.  All the lines with EW errors
larger than 10\% were excluded from the analysis.

(2)~{\it DEIMOS data-set}---
The lower resolution of DEIMOS causes a high degree of line blending and blanketing in the
observed spectra. The
derivation of the abundances through the method of the EWs is therefore
quite uncertain. 
Thus, the iron abundances for this dataset have been measured by  comparing the observed spectra
with a grid of synthetic spectra,  according to the procedure described in \citet{m12_2419}. Each
Fe~I line has been  analyzed individually by performing a $\chi^2$-minimization between the
normalized observed spectrum and  a grid of synthetic spectra. The synthetic spectra have been computed with the
code SYNTHE (\citealt{sbordone04}) assuming the proper atmospheric parameters for each
individual star, then convolved at the DEIMOS resolution and finally resampled at the pixel size of the observed spectra.
To improve the quality of the fit, the normalization is iteratively readjusted locally
in a region of $\sim$50-60 \AA{}.  We estimated the uncertainties in the fitting procedure for each
spectral line by using  Monte Carlo simulations: for each line, Poissonian noise is added to the
best-fit synthetic spectrum in order to reproduce the observed SNR and then the fit is re-computed
as described  above.  The dispersion of the abundance distribution derived from 1000 Monte Carlo
realizations has been  adopted as the abundance uncertainty (typically about $\pm$0.2
dex). 

\section{Error budget}\label{caveats}

In order to verify the robustness of our abundance analysis, in the following we discuss
the effect of each specific assumption we made and the global uncertainty on the
iron abundance estimates.

\subsection{Systematic effects}

\begin{enumerate}
\item {\it Choice of the isochrone.}
The atmospheric parameters of the selected targets have been determined from the projection
onto an isochrone corresponding to the old, sub-solar population 
(see Fig.\ref{cmds}). However, as discussed by F09 and O11, Terzan~5 hosts at least two stellar populations with different 
iron abundances and possibly ages. 
In order to quantify the effect of using isochrones with different metallicity/age, 
we re-derived the atmospheric parameters by using a BaSTI isochrone 
\citep{basti} with an age of 6 Gyr, Z=+0.03 and a solar-scaled mixture (corresponding to [Fe/H]=+0.26 dex). 
The temperatures of the targets decrease by less than 200 K and the gravities 
increase by $\sim$0.2 (as a consequence of the larger evolutive mass). By re-analyzing the spectra of these stars 
with the new parameters, we obtained very similar iron abundances, the mean difference and rms scatter being
$\langle$[Fe/H]$_{{\rm 6~Gyr}}$-[Fe/H]$_{{\rm 12~Gyr}}\rangle=0.00$ dex and $\sigma=0.12$ dex, respectively.
We performed an additional check by adopting the metallicity of the extreme metal poor component  
([Fe/H]$\simeq-0.8$ dex), by using 
a BaSTI isochrone with an age of 12 Gyr, Z$=0.004$ and $\alpha$-enhanced (corresponding to [Fe/H]$\simeq-1$ dex), 
finding that iron abundances increase only by about 0.06 dex.

\item {\it Temperature scale.} To check the impact of different T$_{{\rm eff}}$ scales we derived the 
atmospheric parameters by adopting the Dartmouth \citep{dotter} and Padua \citep{marigo} isochrones, 
and we found negligible variations ($\delta T_{{\rm eff}}\le$50 K). 
Also the adoption of the ($J-K$)--T$_{eff}$ empirical scale by \citet{m98}
has a marginal impact (smaller than 100 K) on the derived temperatures. Such differences lead to iron 
variations smaller than 0.05 dex.

\item {\it Microturbulent velocities.} The assumption of a different 
value of $v_{turb}$ has the effect of shifting the metallicity distribution, without 
changing its shape. Typically, a variation of $\pm$0.1 km$\,$s$^{-1}$ 
leads to iron abundance variations of $\mp$0.07-0.1 dex. Given the typical dispersion of $v_{turb}$
for this kind of stars (see M14a), this effect would lead to a systematic shift of 
the distribution of a few tenths of dex. However the nice match between the abundances measured in these work and 
those obtained by O11 and O13 from higher-resolution spectra for the targets in common 
(see Section \ref{sample}) demonstrates that our choice of $v_{turb}$ is adequate.

\item {\it Model atmospheres.} We repeated the analysis of the targets by adopting MARCS \citep{gustaf} 
and ATLAS9-APOGEE \citep{mesz}
model atmospheres, instead of the ATLAS9 models by \citet{atlas}. The adoption of different model atmospheres 
calculated assuming different lists for opacity, atomic data and computation recipes leads to 
variations smaller than $\pm$0.1 dex in the [Fe/H] determination, 
and it does not change the shape of the metallicity distribution.

\end{enumerate}

\subsection{Abundance uncertainties}\label{uncert}

As discussed in M14a, the global uncertainty of the derived iron abundances (typically
$\sim0.2$ dex) has been computed as the sum in quadrature of two different
sources of error.

{\it (i)}  The first one is the error arising from the
uncertainties on the atmospheric parameters.
Since they have been derived from photometry, the formal uncertainty on these quantities
depends on all those parameters which can affect the location of the
targets in the CMD, such as photometric errors ($\sigma_{{\rm K}}$ and $\sigma_{{\rm J-K}}$
for the magnitude and the color, respectively), 
uncertainty on the absolute and differential reddening ($\sigma_{{\rm[E(B-V)]}}$ and 
$\sigma_{{\rm \delta[E(B-V)]}}$, respectively) and errors on the distance modulus
($\sigma_{{\rm DM}}$). In order to evaluate the uncertainties on $T_{{\rm eff}}$ and
$\log$~$g$ we therefore repeated the projection onto the isochrone for every single
target assuming $\sigma_{K}=0.04$, $\sigma_{J-K}=0.05$, $\sigma_{\delta[E(B-V)]}=0.05$
for the targets in the ACS sample (\citealt{massari}),
$\sigma_{\delta[E(B-V)]}=0.1$ for targets in the WFI FoV (M14a), and
$\sigma_{[E(B-V)]}=0.05$ and $\sigma_{DM}=0.05$ (\citealt{valenti10}). 
We found that uncertainties on T$_{{\rm eff}}$
range from $\sim$60 K up to $\sim$120 K, and those on $\log$~$g$ are of the order of 0.1-0.15 dex. 
For $v_{turb}$ we adopted a conservative uncertainty of 0.2 km$\,$s$^{-1}$.

{\it (ii)} The second source of error is the internal abundance 
uncertainty. For each target this was estimated as the dispersion
of the abundances derived from the lines used, divided
by the squared root of the number of lines. It is worth noticing that, for
any given star, the dispersion is calculated by weighting the abundance
of each line by its own uncertainty (as estimated by DAOSPEC for the
FLAMES targets, and from Monte Carlo simulations for the DEIMOS
targets).

\section{RESULTS}\label{results}

\subsection{Metallicity distribution}\label{sample}

In order to build the metallicity distribution of Terzan 5, we selected bona fide
members according to the following criteria:

{\it (i)}~we considered only stars within the tidal radius of Terzan~5 ($\sim4.6$\arcmin, \citealt{l10}, see also \citealt{miocchi13});

{\it (ii)}~we considered stars with radial velocities within $\pm 2.5\sigma$
(between $-123$ km$\,$s$^{-1}$ and $-43$ km$\,$s$^{-1}$) 
around the systemic radial velocity of
Terzan 5 (v$_{{\rm rad}}\simeq-83$ km$\,$s$^{-1}$, Ferraro et al. 2014 in preparation);

{\it (iii)}~we discarded spectra affected by TiO molecular bands, which can make difficult the
evaluation of the continuum level and in the most extreme cases they completely hide the
spectral lines of interest. To evaluate the impact of TiO bands on the observed spectra we 
followed the strategy described in M14a, adopting the same q-parameter 
(defined as the ratio between the deepest feature of the TiO band at
$\sim8860$ \AA{} and the continuum level measured in the adjacent spectral range
$8850$\AA{}$<\lambda<8856$ \AA{}). 
Thus we analyzed the full set of absorption lines in all the targets
with q$>0.8$, while we adopted a reduced linelist (by selecting only iron absorption 
lines in the range $8680$ \AA{}$<\lambda<8850$ \AA{}, which are only marginally affected by
TiO contamination) for stars with $0.6<$q$<0.8$, and we
completely discarded all the targets with q$<0.6$ (see the empty symbols in Figure \ref{cmds}).

Following these criteria, we selected a sample of 224 stars (170 from the FLAMES dataset and 54
from the DEIMOS dataset). A few stars observed with different instruments were used to check the
internal consistency of the measures.   
In fact, three DEIMOS targets are in common with the FLAMES sample and the average difference between the
metallicity estimates is  $\langle$[Fe/H]$_{{\rm DEIMOS}}-$[Fe/H]$_{{\rm FLAMES}}\rangle$=+0.07$\pm$0.06 ($\sigma$=0.11
dex).  One DEIMOS target is in common with the NIRSPEC sample by O11 and we find 
[Fe/H]$_{{\rm DEIMOS}}$-[Fe/H]$_{{\rm NIRSPEC}}=+0.02$ dex. Finally, three metal-poor FLAMES stars have been
observed at higher spectral resolution with NIRSPEC by O13, and the average difference between
the iron abundance estimates is $0.01 \pm0.02$ dex ($\sigma=0.03$), only. Hence we can conclude that 
iron abundances obtained from different instruments are in good agreement (well within the
errors). For those stars with multiple measurements we adopted the iron abundance obtained from the dataset
observed at higher spectral resolution. Thus the selected sample numbers 220 stars.

\begin{figure}
\plotone{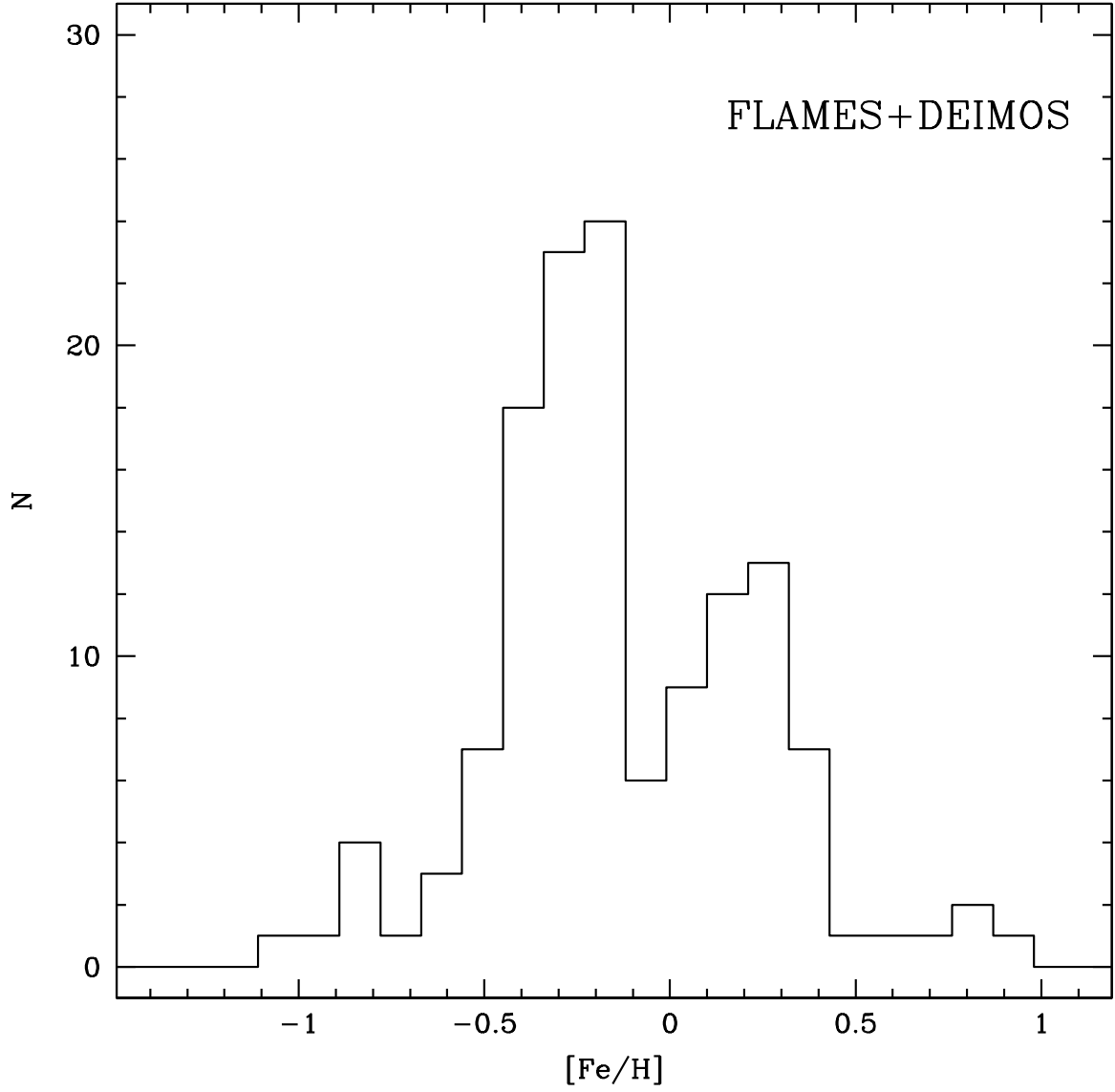}
\caption{\small Metallicity distribution obtained for the unbiased FLAMES+DEIMOS sample (135 targets selected
in the magnitude range $9.6<$K$_c<11.7$), before the statistical decontamination. }
\label{mdf}
\end{figure}

\begin{figure}
\plotone{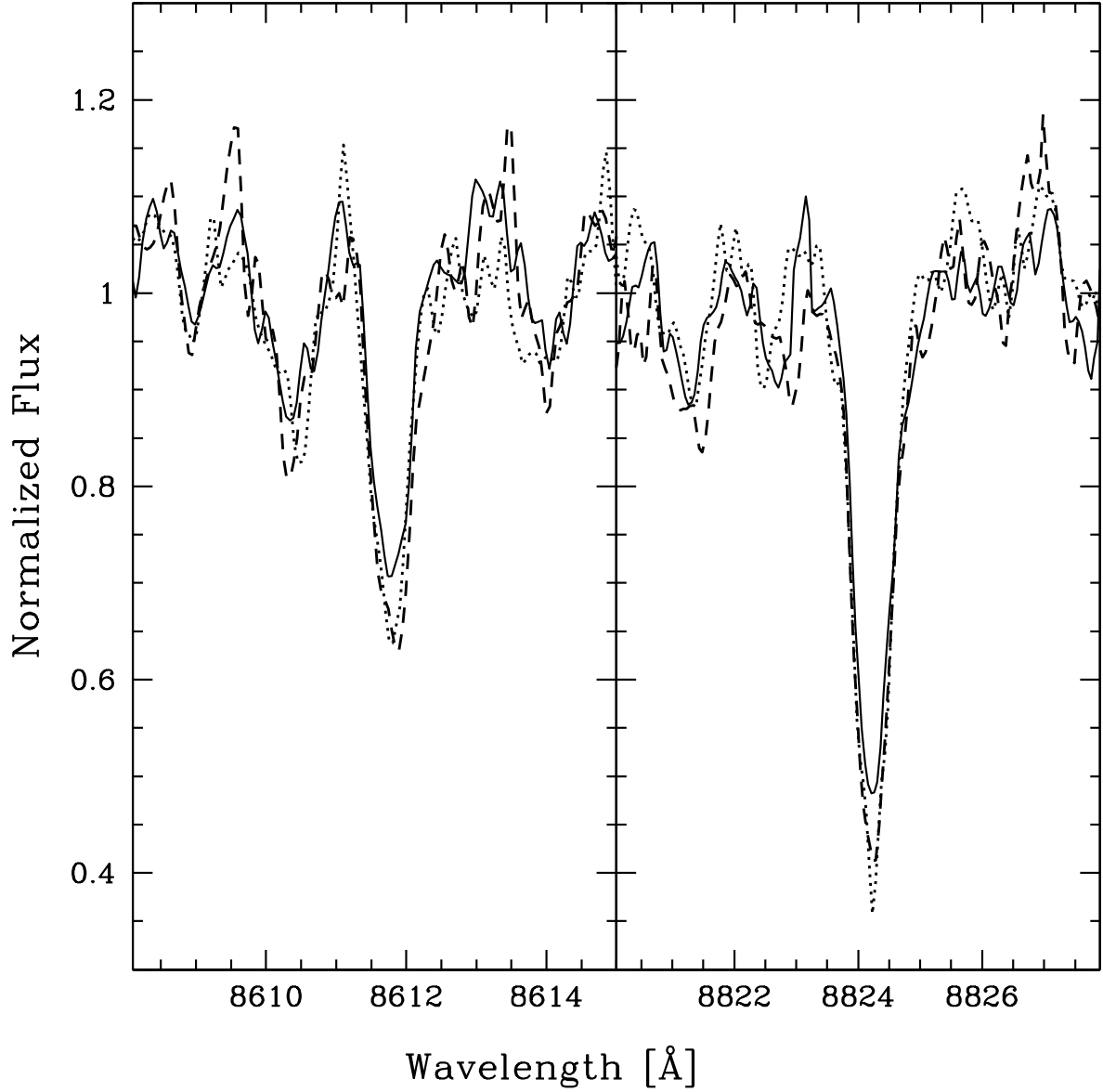}
\caption{\small Comparison of the spectra of two super metal-rich stars 
(namely 7009197 and 7036045, shown as dashed and dotted line, respectively)
and that of a star at [Fe/H]$=+0.26$ (solid line) with similar 
atmospheric parameters.  
The two super metal-rich stars show more pronounced absorption lines, 
thus indicating an actual, very high metallicity.}
\label{supermr}
\end{figure}

As discussed in detail in M14a, the rejection of targets severely contaminated by
TiO bands introduces a bias that leads to the systematic exclusion of metal-rich stars.
To avoid such a bias, we will focus the analysis only on a sub-sample of stars selected
in a magnitude range ($9.6<{\rm K}_c<11.7$) where no targets have been discarded because of 
TiO contamination.
Thus, the final sample discussed in the following contains a total of 135 stars 
and their measured iron abundances and final uncertainties 
(computed as described in Section \ref{uncert}), together with the adopted atmospheric parameters, 
are listed in Table \ref{tab1}. 
The [Fe/H] distribution for these 135 targets is shown in Figure \ref{mdf}.  It is quite
broad, extending from [Fe/H]=--1.01 to +0.94 dex, with an average value of [Fe/H]$=-0.12$ and a dispersion
$\sigma=0.35$, much larger than the typical uncertainty on the abundance estimates. 
More in details, the observed distribution shows a main peak at [Fe/H]$\sim-0.30$ dex
and a secondary component at [Fe/H]$\sim+0.30$ dex, in very good agreement with the results of O11.
Also the third component discovered by O13 is clearly visible at [Fe/H]$\simeq-0.8$ dex.
The distribution also shows a very metal-rich tail, up to [Fe/H]$\sim+0.8$ dex. However,
only five stars have been measured with such an extreme metallicity value, with a somewhat 
larger uncertanty ($\sim 0.2$ dex).  
Figure \ref{supermr} shows the spectra of two such super metal-rich stars 
(7009197 and 7036045 with metallicity of [Fe/H]$=+0.77$ dex and [Fe/H]$=+0.74$ dex, respectively), 
and the spectrum of a star with [Fe/H]$=+0.26$ dex and very similar 
atmospheric parameters (T$_{{\rm eff}}=4325 K$, $\log$~$g=1.7$ dex for the two super metal-rich 
targets and T$_{{\rm eff}}=4269 K$ and $\log$~$g=1.6$ dex for the latter). As can be seen,
the super metal-rich stars have deeper iron absorption lines, thus
indicating a higher metal content with respect to the star at [Fe/H]$=+0.26$ dex. Note that
in order to fit these lines with an iron abundance of  0.3 dex, one needs to assume a
significantly warmer ($\sim500$ K)  temperature.
A spectroscopic followup at  higher spectral resolution
is needed to draw a more firm  conclusions about the metal content of these stars.
If their extremely high metallicity were confirmed,
they would be among the most metal-rich stars in the Galaxy.

\subsection{Statistical decontamination}\label{statdeco}

Even though our sample has been selected within the narrow radial velocity range
around the systemic velocity of Terzan 5, we may expect some
contamination by a few bulge field stars. 
Hence, we performed a statistical decontamination by using the properties of the field
population surrounding Terzan 5 described in M14a.
As shown in detail in that paper, we found that the bulge field population has a very broad radial velocity 
distribution, peaking at v$_{{\rm rad,field}}\sim21$ km$\,$s$^{-1}$ and with a dispersion $\sigma\sim113$ 
km$\,$s$^{-1}$, thus overlapping the Terzan~5 distribution.
When considering different metallicity bins, the bulge population is distributed as follows:
($i$)~$3$\% with [Fe/H]$<-0.5$ dex; ($ii$)~$44$\% with $-0.5<$[Fe/H]$<0$ dex; 
($iii$)~$49$\% with $0<$[Fe/H]$<0.5$ dex; ($iv$)~$4$\% with [Fe/H]$>0.5$ dex.

\begin{figure}
\plotone{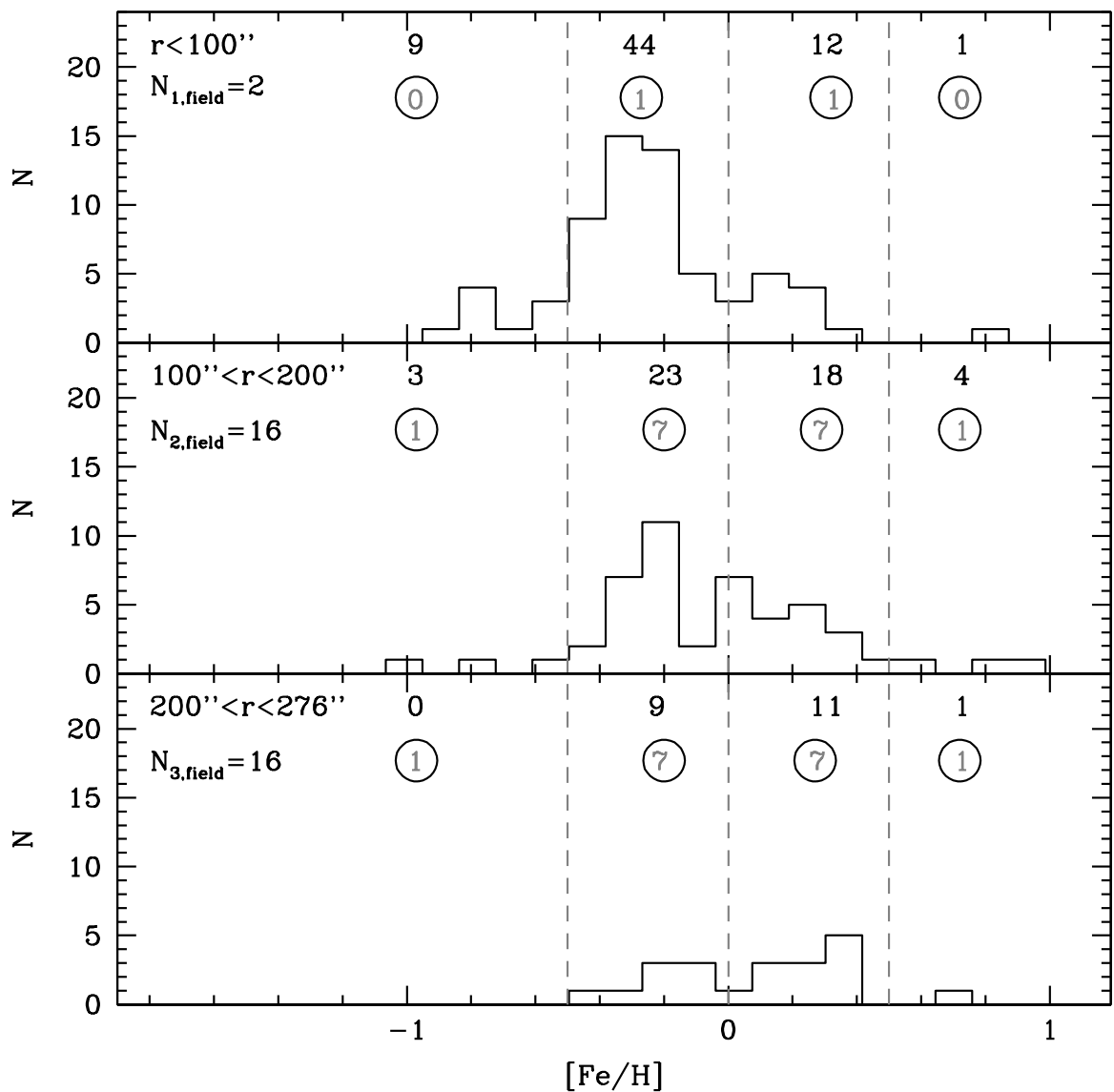}
\caption{\small Metallicity distributions of Terzan 5 stars in the   
inner r$<100$\arcsec ({\it upper panel}), intermediate $100$\arcsec$<$r$<200$\arcsec ({\it middle panel})  
and outer $170$\arcsec$<$r$<276$\arcsec ({\it lower panel}) annuli. The total
number of expected contaminants in each radial bin is reported in the upper-left corner of each panel.
The number of stars observed in each metallicity bin (delimited by vertical dashed lines)
is quoted, while the number of contaminants to be statistically subtracted is highlighted 
in grey and encircled in black. }
\label{deco}
\end{figure}

To perform a meaningful statistical decontamination we first split our sample 
in three radially selected sub-samples (see Fig. \ref{deco}). 
The inner (r$<100$\arcsec) subsample is composed of 66 stars. 
The fractions of field stars expected (Ferraro et al. 2014) to populate this inner region amounts to 2\%,
corresponding to a number of contaminating targets of about N$_{1,field}=2$.
The intermediate subsample ($100$\arcsec$<$r$<200$\arcsec) is composed of 48 stars. 
In this case, the number of expected field stars increases to N$_{2,field}$=16, 
i.e. the 32\% of the subsample.
Finally, in the outer sample ($200$\arcsec$<$r$<276$\arcsec), where we count 21 stars, the expected
contamination by non-member stars amounts to 73\% (corresponding 
to N$_{3,field}=16$).
Fig.~\ref{deco} summarizes the number of stars observed (in black) and the number of contaminants
expected (in grey, encircled) in each radial and metallicity bin considered.

For each radially selected sub-sample and metallicity bin, we then randomly subtracted the corresponding
number of expected contaminants, thus obtaining the decontaminated sample.

\subsection{Decontaminated distribution}\label{gmm}

\begin{figure}
\plotone{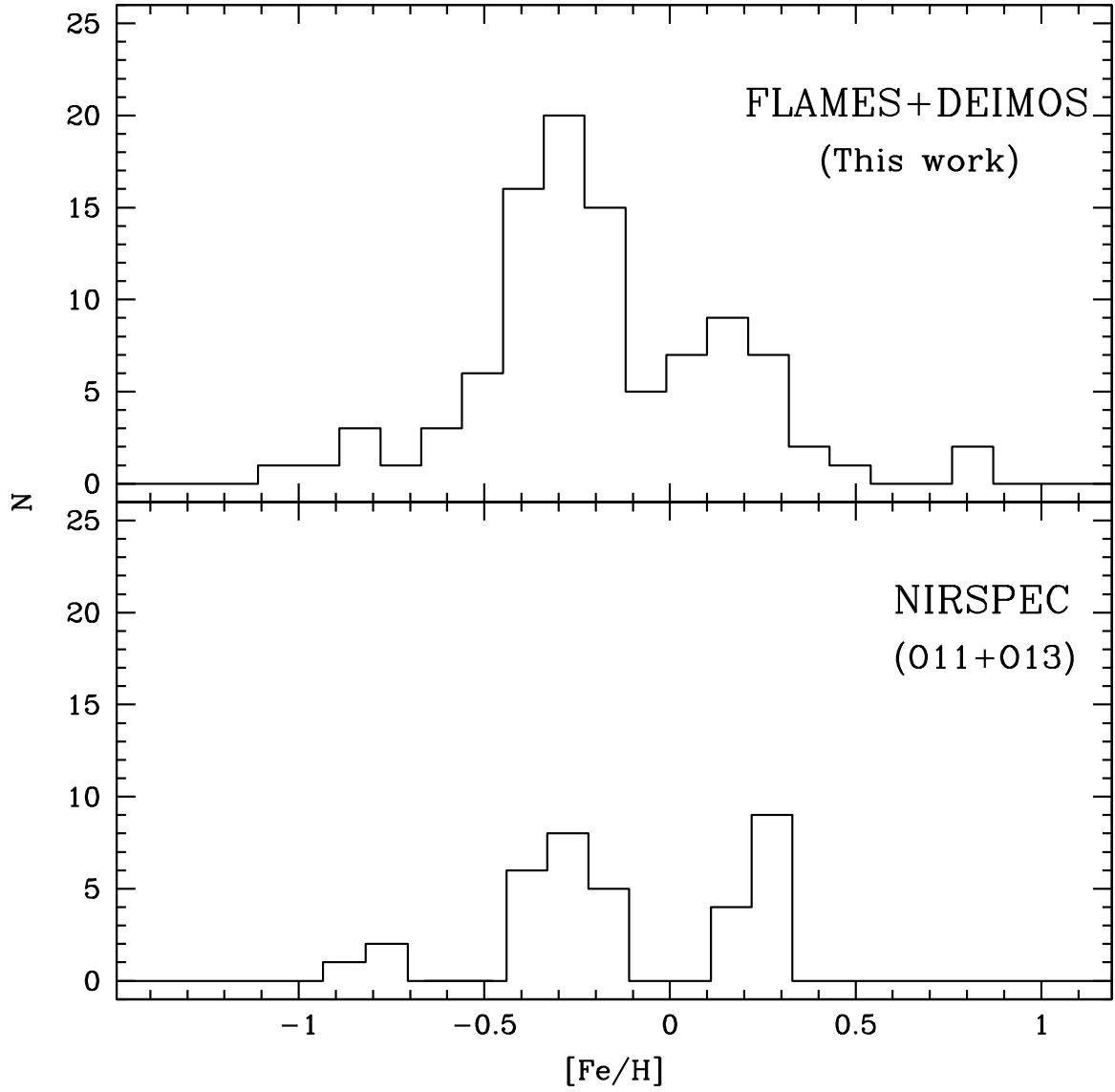}
\caption{\small Statistically decontaminated metallicity distribution for the FLAMES+DEIMOS sample (101 stars, upper panel), compared 
to that derived by O11 and O13 (34 and 3 stars respectively, lower panel). }
\label{mdfdeco}
\end{figure}

The final decontaminated sample is composed of 101 stars and its metallicity
distribution is shown in the upper panel of Fig.~\ref{mdfdeco}. For
comparison, the lower panel shows the distribution of the 34 giant stars in the 
innermost region (r$< 22\arcsec$) of Terzan~5 analyzed in O11 and the three 
metal-poor stars studied in O13. 
The two main peaks at sub-solar and super-solar metallicity, as well as the peak of the minor (5\%) metal-poor component 
at [Fe/H]$\sim$-0.8 dex nicely match each other in the two distributions.

It is worth noticing that, while in the O11 sample the super-solar component is about as numerous as the sub-solar one 
(40\% and 60\%, respectively), in the FLAMES+DEIMOS 
distribution the component at $\sim-0.3$ dex is dominant. This essentially reflects the different radial 
distributions of the two stellar populations observed in Terzan~5, with the metal-rich
stars being more concentrated (at $r<20\arcsec$), 
and rapidly vanishing at $r\gsim50\arcsec$ (see F09 and \citealp{l10}). Note, in fact, that while the
34 RGB stars observed by O11 are located at $r<22\arcsec$, almost all the FLAMES+DEIMOS 
targets are at larger radial distances.
In the FLAMES+DEIMOS distribution there are also three stars with very high metallicities ([Fe/H]$>+0.7$ dex). 
Given the small number of objects, at the moment we conservatively do not consider it as an additional sub-population of Terzan 5. 

\begin{figure}
\plotone{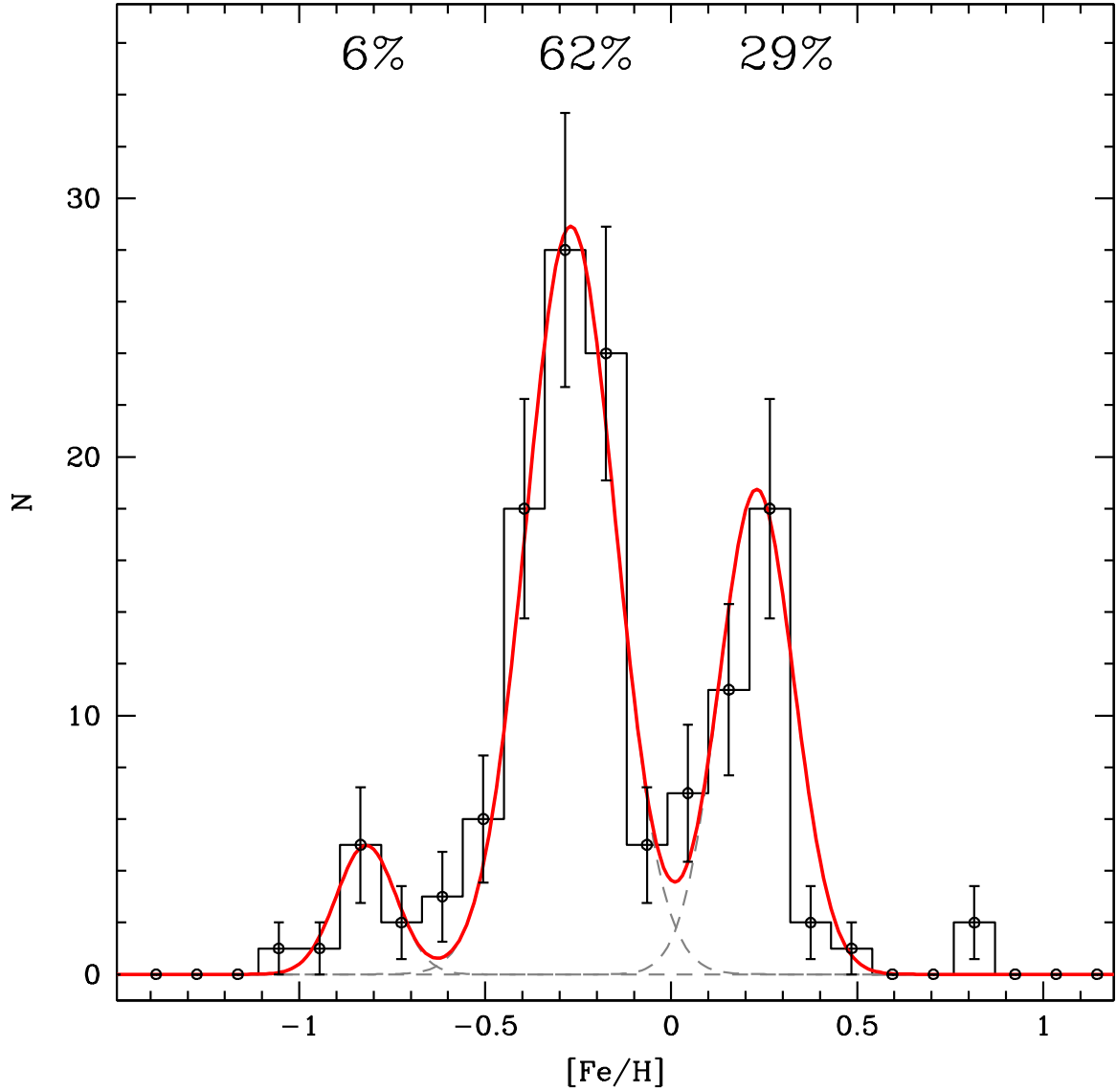}
\caption{\small Decontaminated metallicity distribution for the combined FLAMES+DEIMOS (101 stars, this work) and NIRSPEC (34 targets, O11) spectroscopic samples. 
The solid red line shows the fit that best reproduces the observed distribution using three Gaussian profiles. Individual Gaussian 
components are shown as grey dashed lines. The percentage of each individual component with respect to the total sample
of 135 stars is also reported.}
\label{mdf2}
\end{figure}

The overall metallicity distribution of Terzan~5, derived from a total of $135$ stars
(corresponding to $101$ targets from the decontaminated FLAMES+DEIMOS sample discussed here, plus 
$34$ NIRSPEC giants from O11) is shown in Fig.~\ref{mdf2}.  
In order to statistically verify the apparent multi-modal behavior of the
distribution, we used the Gaussian mixture modeling (GMM) algorithm proposed by 
\cite{gmm}. This algorithm determines whether a distribution is better
described by a unimodal or a bimodal Gaussian fit. In particular, three requirements
are needed to rule out the unimodality of a distribution:
\begin{enumerate}
 \item the separation D between the peaks, normalized to the widths of the 
 Gaussians, defined as in \cite{ashman94}, has to be strictly 
 larger than 2;
\item the kurtosis of the distribution has to be negative;
\item the likelihood ratio test (\citealt{wolfe}), which obeys
$\chi^{2}$ statistics, has to give sufficiently large values of $\chi^{2}$.
\end{enumerate}
The algorithm also performs a parametric bootstrap to determine the confidence 
level at which the unimodality hypothesis can be accepted or rejected.

First of all, we computed the GMM test on the two main components.
In this case, all the three requirements are verified (D$=3.96$, kurtosis$=-0.89$
and $\chi^{2}=43.46$ with 4 degrees of freedom) and the unimodal fit is rejected 
with a probability P$>99.9$\%.
We then repeated the same procedure considering the most metal-poor component
at [Fe/H]$\simeq-0.8$ and the sub-solar one. 
Also in this case the unimodal fit is rejected with a probability P$>99.9$\%\footnote{
Note that because of the large difference in size between the two components, the computed
kurtosis turns out to be positive. However we checked that by reducing the size of the sample 
belonging to the sub-solar component, the kurtosis turns negative, as required by the GMM test.}.
We can therefore conclude that the metallicity distribution of Terzan 5 
is clearly \emph{multi-peaked}. 
We are able to reproduce its shape using three Gaussian profiles (red line in Fig. \ref{mdf2}).
Adopting the mean values and dispersions obtained from the GMM test, the two main 
peaks are located at [Fe/H]$\simeq -0.27$ dex (with $\sigma=0.12$) and [Fe/H]$\simeq +0.25$
dex (with $\sigma=0.12$), the sub-solar component being largely dominant 
(62\% of the total). A minor component (6\% of the total) is
located at [Fe/H]$\simeq-0.77$ dex (with $\sigma=0.11$).

\begin{figure}
\plotone{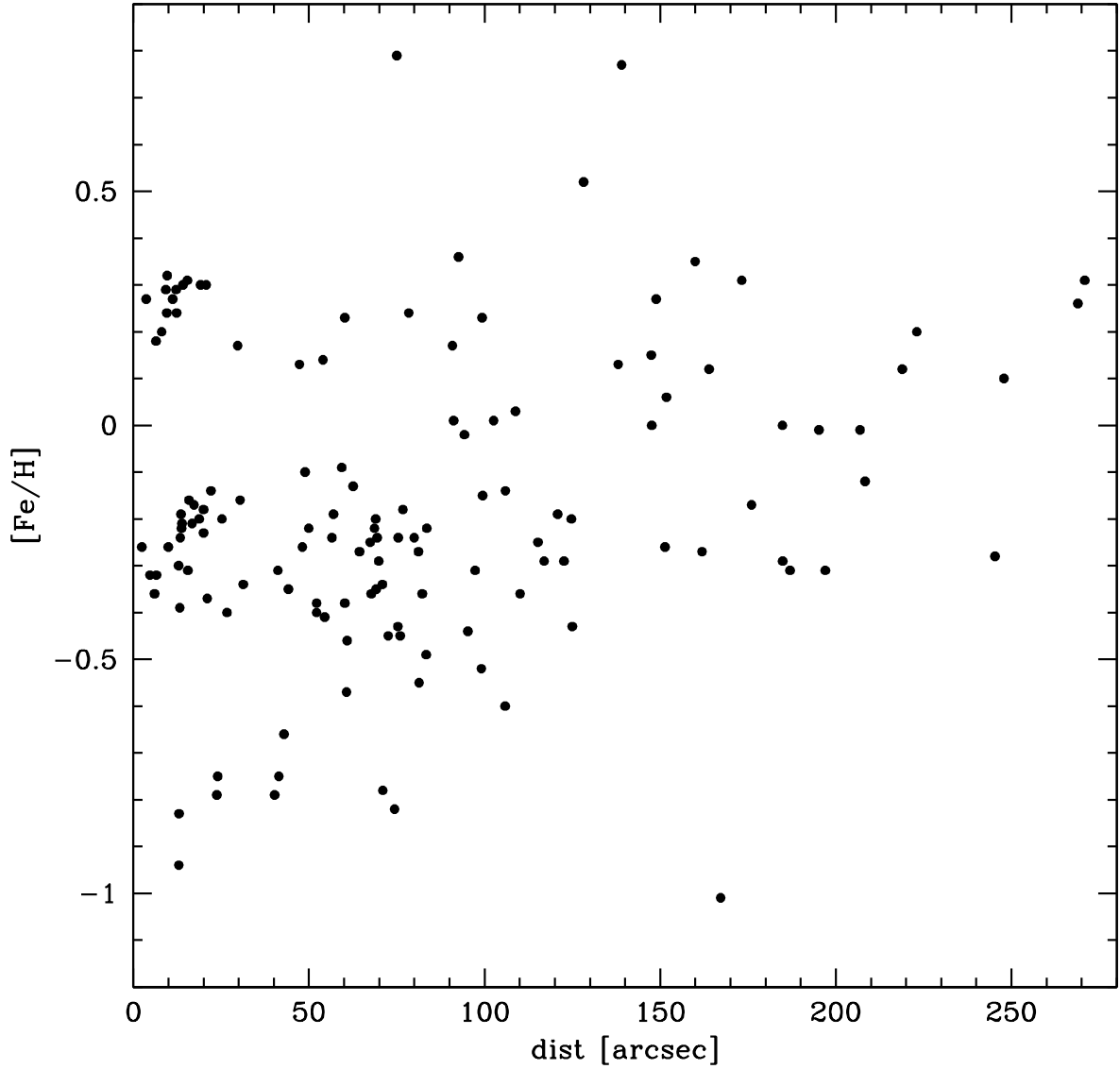}
\caption{\small  The [Fe/H] distribution as a function of the distance from the cluster center 
for the 135 stars composing the final decontaminated iron distribution shown in Fig \ref{mdf2}: 
the multi-modality of the metallicity distribution is clearly evident. The bulk of 
each of the three components
is located in the innermost $80$\arcsec~  from the cluster center,
 thus further confirming the actual membership of all the three populations.}
\label{distvsfe}
\end{figure}

Finally in  Fig. \ref{distvsfe}  we show the radial distribution of the 135 stars (101 from this study
and 34 from O11) adopted 
to construct the Terzan 5 metallicity distribution shown in Fig. \ref{mdf2}:
the multi-modal metallicity distribution is clearly evident also in this plot. It is worth 
of noticing that the most metal poor component is essentially located in the innermost 
$80$\arcsec~ from the cluster center, further supporting the membership of this minor component.

\section{DISCUSSION AND CONCLUSIONS}\label{discuss}

The results presented in this work are based on a statistically significant sample of stars 
distributed over the entire radial extent of Terzan 5, thus solidly
sampling the metallicity distribution of this stellar system.
We confirm the previous claims by F09, O11 and O13 
that Terzan~5 hosts multiple stellar populations characterized by significantly different
iron contents.

The multi-modal iron distribution of Terzan~5 puts this stellar system
in a completely different framework with respect to that of genuine GCs.  
In fact, the latter systems, although showing significant spreads in 
the abundance of light elements (as sodium, oxygen,
aluminum etc.; see, e.g., \citealt{carretta10})\footnote{This suggests that
GC formation has been more complex than previously thought, having
re-processed the low-energy ejecta from
asymptotic giant branch stars \citep{ventura01} and/or fast rotating
massive stars \citep{decressin07}, with enrichment timescales of
$\sim10^8$ years or shorter 
\citep[e.g.,][]{dercole,valcarce}.},
still maintain a striking homogeneity in terms of iron content, thus
indicating that their stellar populations formed within a potential
well which was unable to retain the high-velocity gas ejected by violent SN
explosions.  Indeed, the iron content of stellar populations can be
considered the main feature to distinguish between genuine GCs and
more complex stellar systems (\citealt{willman}). Following this view,
Terzan 5 certainly belongs to the latter class of objects.  

Recent high-precision spectroscopic studies have shown some iron spread 
(but still with a range largely smaller than  1 dex) in a few GCs, namely 
M22 \citep{marino09,marino11a,marino12},
M2 \citep{yongm2}, and M54 \citep{m54}\footnote {Other two GCs
  have been proposed to harbor intrinsic iron dispersion, namely NGC
  5824 \citep{saviane12,dacosta14} and NGC~3201 \citep{simmerer13}. We exclude
  these two clusters from our discussion because their intrinsic iron
  scatter has been not firmly confirmed.  The analysis of NGC 5824 is
  based on the Calcium II triplet as a proxy of metallicity and direct
  measurements of iron lines from high-resolution spectra are not
  available yet. Moreover, based on HST photometry, \cite{sanna} have
  recently found that the color distribution of RGB stars is consistent
  with no metallicity spread. Concerning NGC 3201, the analysis of
  \citet{simmerer13} leads to an appreciable iron spread among the stars of
  this cluster, but the analysis of \citet{munoz} contradicts this
  result.}.  
However, the iron distributions observed in these systems are unimodal, 
with no evidence of multiple peaks, as we also verified 
by means of the GMM test described above.  Only M54 shows a tail towards the 
metal-rich side of its metallicity distribution, but this population can 
be severely contaminated by the Sagittarius field stars (see
\citealt{bellazzini99,bellazz08}). 

\begin{figure}
\plotone{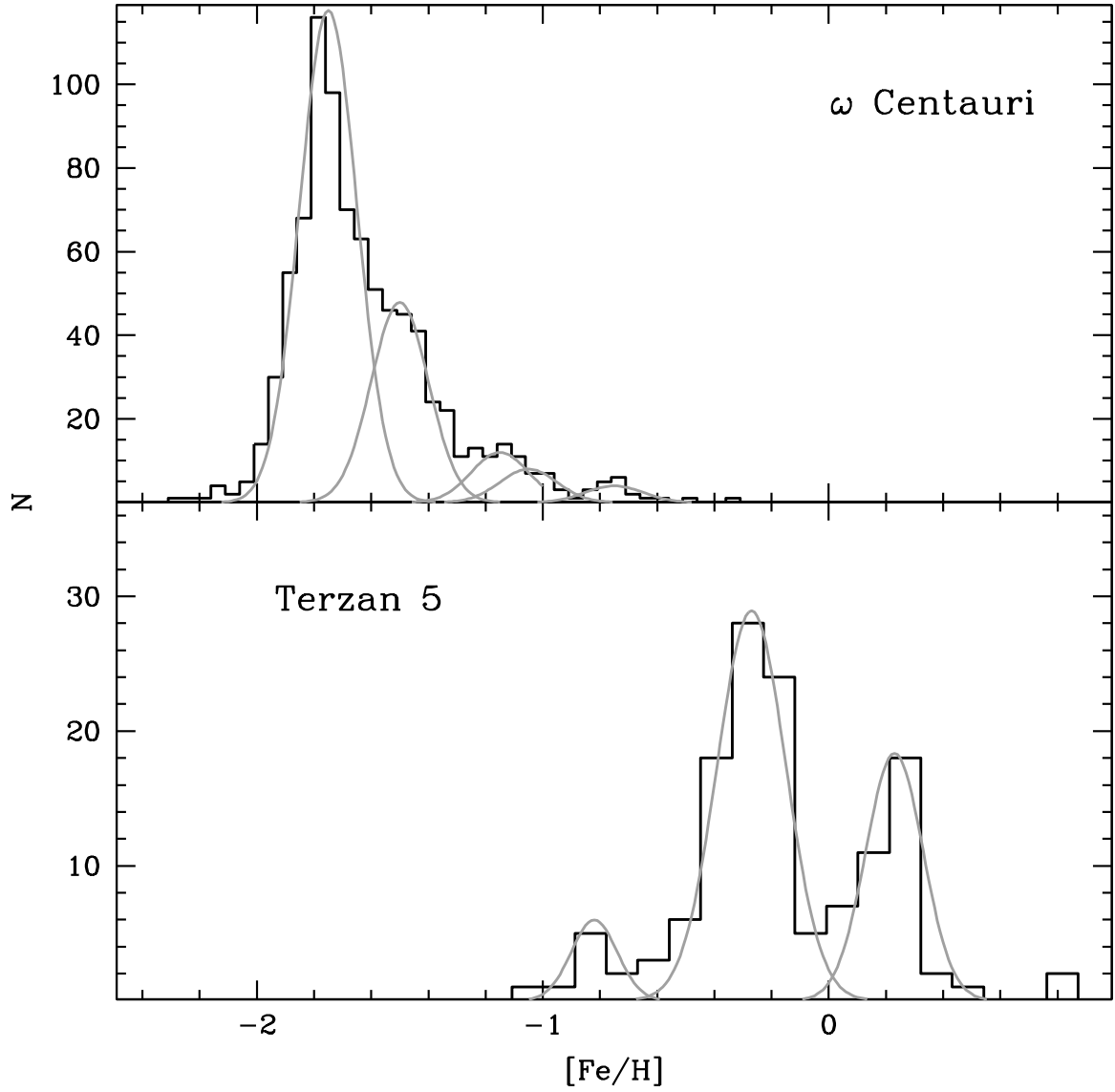}
\caption{\small The metallicity distribution of $\omega$ Centauri ({\it upper panel}) and Terzan 5 ({\it lower panel}).
The distribution of $\omega$ Centauri, together 
with the five Gaussians reproducing its multi-modality, have been taken from \cite{jp10}.}
\label{confr}
\end{figure}

Only another GC-like system in the Galaxy ($\omega$ Centauri) is known
to host a large variety of stellar sub-populations \citep{lee99,
pancino2000, ferraro04, ferraro06, bellini09, bellini10, bellini13} 
with a large range of iron abundance ($\Delta$[Fe/H]$>1$ dex; \citealp{omega,
origlia03, sollima04, sollima07, jp10, villanova14}), similar to what is observed in
Terzan 5.  
As shown in Figure \ref{confr}, a few similarities between
Terzan 5 and $\omega$ Centauri can be indeed recognized: {\it (i)} a
broad extension of the iron distribution ($\sim 1.8$ dex in Terzan 5
and $\sim 2$ dex in $\omega$ Centauri; see \citealt{jp10} for the
latter); {\it (ii)} a multi-modal distribution; 
{\it (iii)} the presence of a numerically small
stellar population ($\sim5-10$\% of the total in both cases) which is more
metal-poor than the main peak, possibly corresponding to the
first generation of stars in the system (see \citealp{pancino11} for
$\omega$ Centauri).
The intrinsic large dispersion in [Fe/H] indicates that in the past
these systems were massive enough to retain the high-energy,
high-velocity ejecta of SNe, allowing for multiple bursts of star
formation from increasingly iron-enriched gas over timescales 
of the order of a few $10^9$ years.

$\omega$ Centauri is now believed
to be the remnant of a dwarf galaxy accreted by the Milky Way
\citep[e.g.,][]{bekki03}. In contrast, the high metallicity regime of Terzan 5 
(not observed in the known satellites of our Galaxy) and its
tight chemical link with the Galactic bulge (O11, O13, M14a) make very unlikely
that it has been accreted from outside the Milky Way, and 
favor an in-situ formation.
Terzan 5 could be the remnant of an early
giant structure which may plausibly have contributed to form the Galactic bulge. 
In principle, the low dispersion of the iron content within each sub-population of Terzan 5 
could be consistent with both a bursty star formation and chemical self-enrichment, and
the dry merging of individual sub-structures with different metallicity \citep[e.g.][]
{immeli04, elmegreen08,fs11}.  
However, the fact that among the three distinct sub-populations, the
metal-rich one is more centrally concentrated than
the more metal-poor ones seems to favor 
a self-enrichment scenario, at least for the formation of the metal-rich component
\citep[e.g.][]{dercole}.

Certainly Terzan 5 is very peculiar, if not unique, system within the Galactic bulge.
In order to solve the puzzle of its true nature, some pieces of information are still missing, such as
the accurate estimate of the absolute ages of its populations, and a
proper characterization of the global kinematical properties of the
system.

\acknowledgements{ We thank the anonymous referee for his/her useful comments and 
  suggestions which helped us to improve the presentation
  of our results. This research is part of the project COSMIC-LAB
  (web site: http://www.cosmic-lab.eu) funded by the European Research
  Council (under contract ERC-2010-AdG-267675). M.B. acknowledges financial 
  support from PRIN MIUR 2010-2011 project “The Chemical and Dynamical 
  Evolution of the Milky Way and Local Group Galaxies”, prot. 2010LY5N2T.”
  R. M. R. acknowledges support from US National Science Foundation grant 
  AST-1212095 and from  grant GO-12933 from the Space Telescope Science Institute.
  Some of the data presented herein were obtained at the W.M. Keck Observatory, 
  which is operated as a scientific partnership among the California Institute of 
  Technology, the University of California and the National Aeronautics and Space Administration. 
  The Observatory was made possible by the generous financial support of the W.M. Keck Foundation. 
  The authors wish to recognize and acknowledge the very significant cultural 
  role and reverence that the summit of Mauna Kea has always had within the 
  indigenous Hawaiian community.  We are most fortunate to have the opportunity 
  to conduct observations from this mountain. 
}

\newpage

\begin{deluxetable}{rrrrrcrrc}
\tablewidth{0pc}
\tablecolumns{8}
\tiny
\tablecaption{Iron abundance of Terzan 5 stars.}
\tablehead{\colhead{ID} & \colhead{RA} & \colhead{Dec} & \colhead{K$_{c}$} & \colhead{T$_{{\rm eff}}$} & \colhead{log~$g$} & \colhead{[Fe/H]} 
& \colhead{$\sigma_{{\rm [Fe/H]}}$} & Dataset \\
& & & \colhead{(mag)} & \colhead{(K)} & \colhead{(cm\,s$^{-2}$)} & \colhead{(dex)} & \colhead{(dex)} & }
\startdata
 & & \\
        109  &    266.9801977   &    -24.7835577  &  8.60  &   3741   &   0.7  &  --0.30   &    0.17   &     FLAMES \\
        126  &    267.0292394   &    -24.7803417  &  8.60  &   3736   &   0.7  &  --0.26   &    0.14   &     FLAMES \\
        134  &    267.0332227   &    -24.7953548  &  8.73  &   3771   &   0.7  &  --0.32   &    0.07   &     FLAMES \\
        146  &    267.0254477   &    -24.7817867  &  8.78  &   3786   &   0.8  &  --0.38   &    0.07   &     FLAMES \\
        148  &    267.0291700   &    -24.7969272  &  8.86  &   3804   &   0.8  &  --0.17   &    0.06   &     FLAMES \\
        155  &    267.0286940   &    -24.7786346  &  8.83  &   3799   &   0.8  &  --0.34   &    0.09   &     FLAMES \\
        158  &    267.0124475   &    -24.7843182  &  8.88  &   3814   &   0.8  &  --0.36   &    0.10   &     FLAMES \\
        159  &    267.0226507   &    -24.7624999  &  8.83  &   3800   &   0.8  &  --0.32   &    0.06   &     FLAMES \\
        164  &    267.0282685   &    -24.7949808  &  8.98  &   3838   &   0.8  &  --0.31   &    0.07   &     FLAMES \\
        165  &    267.0315361   &    -24.7896355  &  8.95  &   3831   &   0.8  &  --0.18   &    0.15   &     FLAMES \\
\enddata
\tablecomments{\small Identification number, coordinates, K$_c$ magnitude atmospheric parameters, iron abundances and
their uncertainties, and corresponding dataset for  all the 220 stars members of Terzan 5 with iron 
abundance measured. All stars with K$_{c}<9.6$ or K$_{c}>11.7$ have been excluded from the analysis of the
MDF (see Section \ref{sample}).}
\label{tab1}
\end{deluxetable}

\end{document}